# Superconductivity in iron silicide $Lu_2Fe_3Si_5$ probed by radiation-induced disordering

A. E. Karkin, M. R. Yangirov, Yu. N. Akshentsev and B. N. Goshchitskii

*Institute of Metal Physics UB RAS, 18 S. Kovalevskoi Str., Ekaterinburg 620219, Russia*

e-mail: aekarkin@rambler.ru

Resistivity $\rho(T)$, Hall coefficient $R_H(T)$, superconducting temperature $T_c$, and the slope of the upper critical field $-dH_{c2}/dT$ were studied in poly- and single-crystalline samples of the Fe-based superconductor $Lu_2Fe_3Si_5$ irradiated by fast neutrons. Atomic disordering induced by the neutron irradiation leads to a fast suppression of $T_c$ similarly to the case of doping of $Lu_2Fe_3Si_5$ with magnetic (Dy) and non-magnetic (Sc, Y) impurities. The same effect was observed in a novel FeAs-based superconductor La(O-F)FeAs after irradiation. Such behavior is accounted for by strong pair breaking that is traceable to scattering at non-magnetic impurities or radiation defects in unconventional superconductors. In such superconductors the sign of the order parameter changes between the different Fermi sheets ($s^{\pm}$ model). Some relations that are specified for the properties of the normal and superconducting states in high-temperature superconductors are also observed in $Lu_2Fe_3Si_5$. The first is the relationship $-dH_{c2}/dT \sim T_c$, instead of the one expected for dirty superconductors $-dH_{c2}/dT \sim \rho_0$. The second is a correlation between the low-temperature linear coefficient $a$ in the resistivity $\rho = \rho_0 + a_1 T$, which appears presumably due to the scattering at magnetic fluctuations, and $T_c$; this correlation being an evidence of a tight relation between the superconductivity and magnetism. The data point to an unconventional (non-fononic) mechanism of superconductivity in $Lu_2Fe_3Si_5$, and, probably, in some other Fe-based compounds, which can be fruitfully studied via the radiation-induced disordering.

The discovery of high-temperature superconductivity in layered iron-based compounds [1] stimulated active experimental and theoretical studies of these systems in view of the possibility of the Cooper pairing of charge carriers by an anomalous type. Hence, a systematic study of the disordering effects in new superconductors is especially important [2]. According to the Anderson theorem [3], nonmagnetic impurities do not cause a suppression of the superconductivity in the case of a conventional $s$-type isotropic pairing. If the singlet pairing is traceable to the exchange of spin excitations, the requirement for this is the symmetry with a sign-changing order parameter [4]. Evidently, such requirement is fulfilled in high-$T_c$ cuprates, where pairing with $d$-wave symmetry is realized, while the pairing process proper is destroyed by an intraband scattering at nonmagnetic centers [4, 5, 6]. In the FeAs-based superconductors, the ordering parameter has the $s$-type symmetry, therefore a generally accepted is the $s^{\pm}$-model, which treats a superconducting state with the opposite signs of the ordering parameter for electrons and holes [7, 4, 8, 9]. In this case nonmagnetic scatters must lead to the suppression of superconductivity due to the interband scattering between the electron- and hole-type Fermi surfaces [4, 5, 10]. Thus, the study of the atomic disordering in superconducting systems in which\ nonmagnetic scatters are generated allows one to reveal the symmetry of the ordering parameter.

Fast neutron irradiation is the most effective method of atomic disordering which was successfully applied earlier in investigation of a number of high-temperature superconductors. After irradiation the $MgB_2$ compound demonstrated a relatively weak change of the superconducting temperature $T_c$, which is typical of the systems with a strong electron-phonon interaction and isotropic $s$-type pairing [11, 12, 13]. In the Cu-based superconductors such as $YBa_2Cu_3O_7$ [13, 14, 15, 16, 17, 18] as well as the FeAs-based superconductors [19, 20, 21], the fast and complete suppression of superconductivity under the high-energy particles irradiation evidences a more exotic (non-fononic) pairing mechanism.

In the early 80's, several investigations were carried out to understand the superconductivity exhibited by compounds belonging to the $R_2Fe_3Si_5$ system [22, 23, 24, 25]. These compounds crystallize in a tetragonal structure of the $Sc_2Fe_3Si_5$-type, consisting of a quasi-one-dimensional iron chain along the $c$ axis and quasi-two-dimensional iron squares parallel to the basal plane. In $Lu_2Fe_3Si_5$ the superconductivity occurs at $T_c \sim 6.0$ K which is exceptionally high among the Fe-based compounds other than the FeAs family. Moreover, a remarkable decrease of $T_c$ by nonmagnetic impurities [26, 27, 28] also testify to an unconventional origin of the superconductivity in this compounds.

This paper reports the results of studying the radiation-induced disordering effects on the properties of the superconducting and normal states of $Lu_2Fe_3Si_5$. It was expected that the irradiation defects, as well as impurities such as Sc and Y, would create non-magnetic scattering centers without substantial changes of the band structure. However, substitution of Lu with atoms of the same valences cannot produce a significant disorder (disorder appears in this case as a result of some lattice distortions in the vicinity of substituted sites), while the fast-neutron irradiation allows one to create defects with a much higher scattering ability and, hence, a stronger disorder can be achieved. In the present study our attention is focused on the effect of disordering on $T_c$ and the slope of the upper critical field $-dH_{c2}/dT$, as well as their correlations with the normal-state properties; the disordering-induced changes in the crystal structure being beyond the scope of the work.

Samples of $Lu_2Fe_3Si_5$ were prepared by arc melting stoichiometric amounts of high-purity elements. To improve the homogeneity of polycrystalline samples, they were annealed at 1200°C for 19 hours. Single crystals 1.0×0.2×0.2 mm in size were obtained by annealing of the arc-melted ingot at 1720°C for 2 hours.

The resistivity $\rho$ and Hall coefficient $R_H$ were measured using the standard four-point method with the reverse of the directions of the dc current and magnetic field and switching-over between the current and potential leads [29]. The electric contacts were made by ultrasonic soldering with indium. Measurements were performed in the temperature range $T$ = 1.5 – 380 K in magnetic fields up to 13.6 T. The polycrystalline samples were irradiated with fast neutrons with the fluence $\Phi = 2\cdot10^{19}$ cm$^{-2}$ (for neutron energies $E_n > 0.1$ MeV) at the irradiation temperature $T_{irr}$ = 50 ± 10°C. The single-crystal samples were irradiated with the lower fluence $\Phi = 5\cdot10^{18}$ cm$^{-2}$. The samples of both types were annealed isochronally for 0.5 h in vacuum in the temperature range of $T_{ann}$ = 50 –1000°C with the 50°C step.

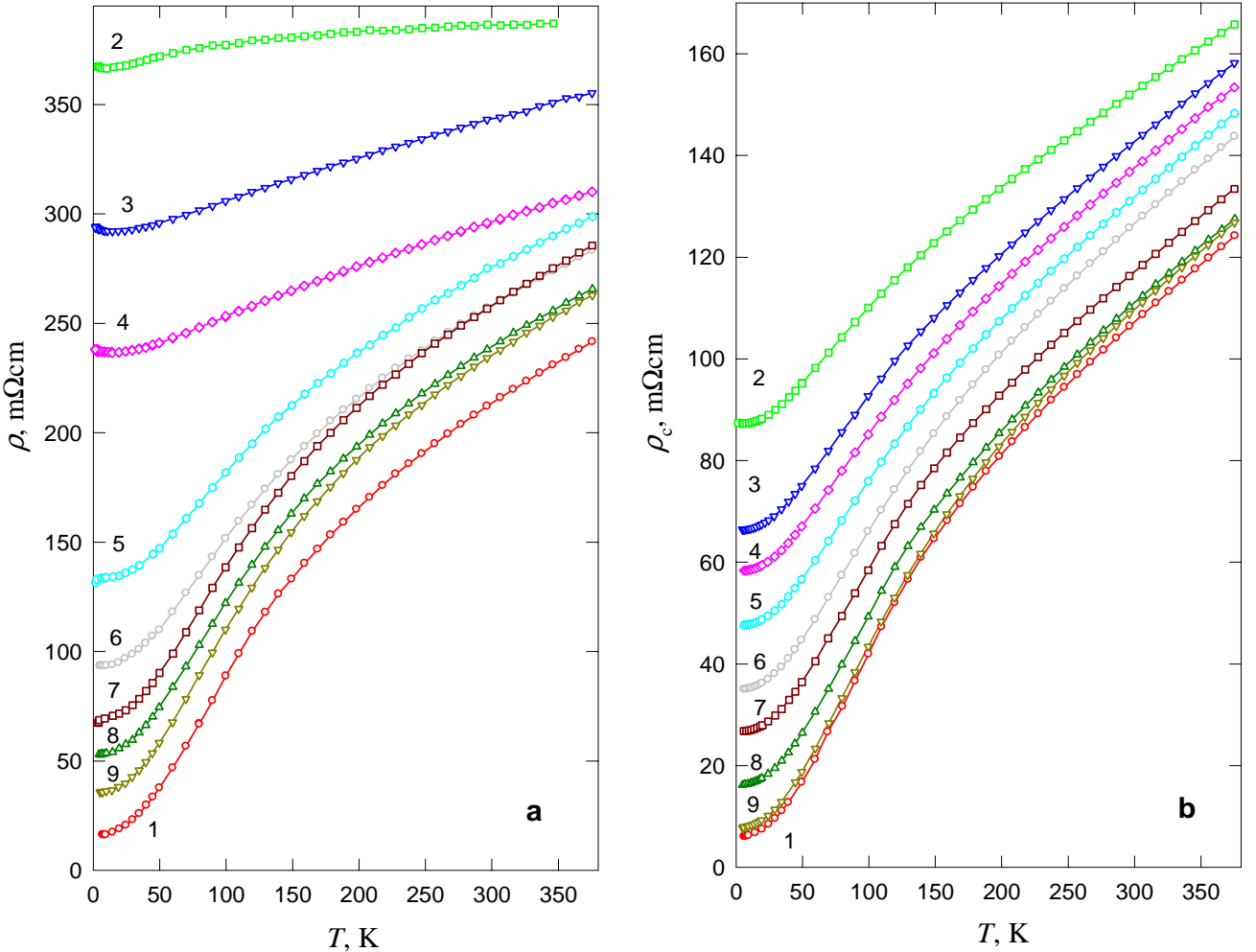

Fig. 1. (color online) (a) Temperature dependences of the normal-state ($T > T_c$) resistivity $\rho$ for the polycrystalline $Lu_2Fe_3Si_5$ sample: initial state (1) and irradiated to the neutron fluence $\Phi = 2\cdot10^{19}$ cm$^{-2}$ and annealed at 350°C (2), 450°C (3), 550°C (4), 750°C (5), 800°C (6), 850°C (7), 950°C (8) and 1000°C (9). (b) Temperature dependences of the resistivity in the c-direction $\rho_c$ for the $Lu_2Fe_3Si_5$ single crystal: initial state (1), irradiated to the neutron fluence $\Phi = 5\cdot10^{18}$ cm$^{-2}$ (2), and annealed at 400°C (3), 500°C (4), 600°C (5), 700°C (6), 750°C (7), 800°C (8) and 900°C (9).

The irradiation to the fast-neutron fluence $\Phi = 2 \cdot 10^{19}$ cm$^{-2}$ (polycrystalline samples) and $\Phi = 5 \cdot 10^{18}$ cm$^{-2}$ (single crystals) suppresses the superconductivity and results in significant changes in the resistivity curves $\rho(T)$. Sequential annealings in the range of $T_{ann}$ = 100 – 1000°C lead to a practically complete restoration of the sample properties in both the normal and superconducting states.

Figures 1a and 1b show the temperature dependences of the resistivity $\rho(T)$ for the polycrystalline and single crystal samples, respectively, representing initial, irradiated, and annealed states. Both sets of data are very similar, taking into account the anisotropy of resistivity $\rho_{ab}/\rho_c \sim 4$ [28], whereas the percolation model [30] predicts that the ratio of the polycrystalline-sample resistivity $\rho$ to the single-crystal resistivity $\rho_c$ must be $\rho/\rho_c \sim 2.7$.

The slope $d\rho/dT$ at high temperatures $T = 100 – 380$ K decreases with increasing $\rho_0$. A similar "saturation" of the resistivity is observed in many strongly disordered metallic compounds, including many compounds irradiated by fast neutrons [13].

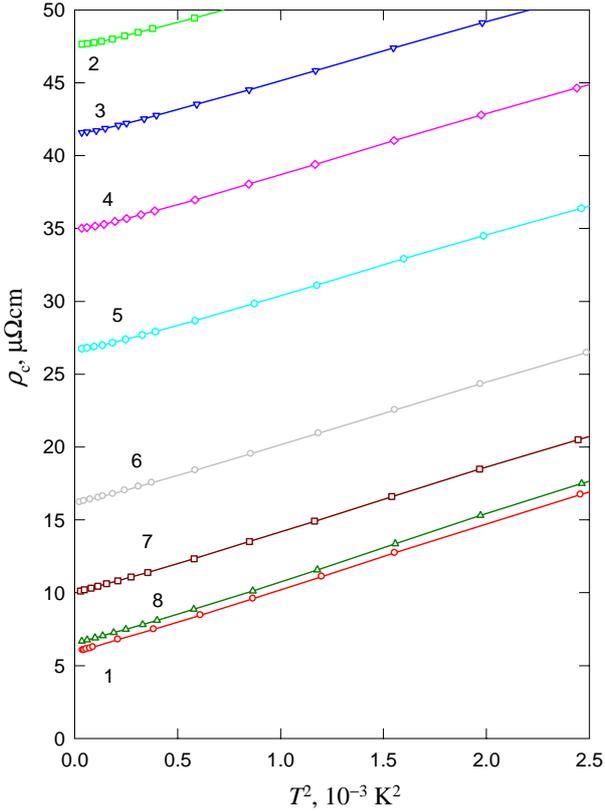

Fig. 2. (color online) Resistivity $\rho$ as a function of $T^2$ for the Lu$_2$Fe$_3$Si$_5$ single crystal: initial state (1) and irradiated to the neutron fluence $\Phi = 5 \cdot 10^{18}$ cm$^{-2}$ and annealed at 600°C (2), 650°C (3), 700°C (4), 750°C (5), 800°C (6), 850°C (7) and 1000°C (8)) Lu$_2$Fe$_3$Si$_5$ single crystal.

In the temperature range $10 < T < 70$ K the $\rho(T)$ curves for the single-crystal sample at low $\rho_0 = 5$-40 µΩcm obey a quadratic law $\rho(T) = \rho_0 + a_2 T^2$ with $a_2 \sim$ $4 \cdot 10^{-3}$ µΩcm/K$^2$ (Fig. 2). The similar behavior is observed for the polycrystalline sample: $a_2$ is approximately constant at $\rho_0 = 5 - 50$ mΩcm and slightly decreases with the further increase in $\rho_0$ (Fig. 1).

At lower temperatures $T < 10$ K the resistivity curves are described better by linear functions $\rho(T) = a_0 + a_1 T$ for the superconducting samples ($\rho_0 < 25$ µΩcm, Fig. 3), while for the non superconducting samples ($\rho_0 > 25$ µΩcm), a small negative slope $d\rho/dT < 0$ is observed.

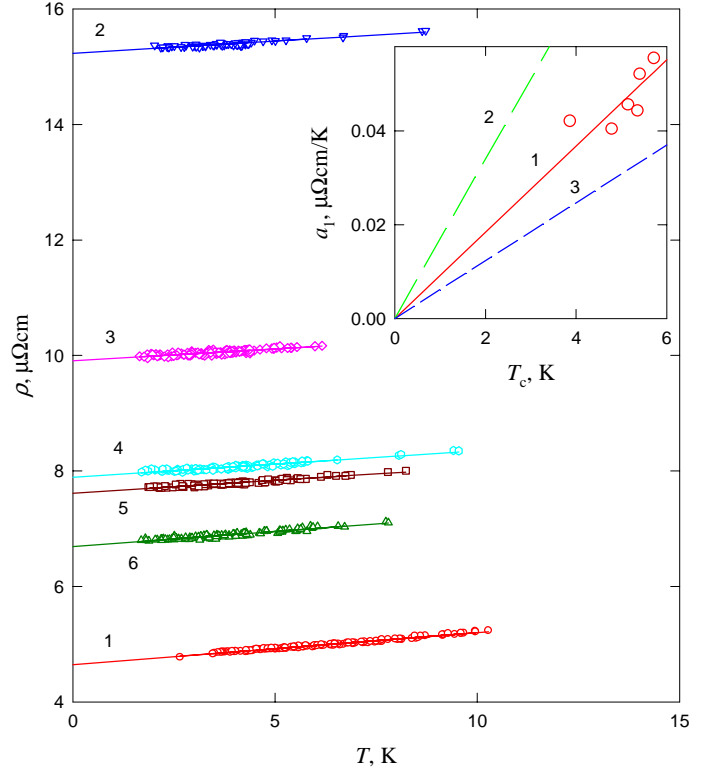

Fig. 3. (color online) Temperature dependences of resistivity $\rho$ for the Lu$_2$Fe$_3$Si$_5$ single crystal: initial state (1) and irradiated to the neutron fluence $\Phi = 5 \cdot 10^{18}$ cm$^{-2}$ and annealed at 800°C (2), 850°C (3), 900°C (4), 950°C (5) and 1000°C (6). The points are collected from the curves at $T > T_c$ in magnetic fields up to 13.6 T with the correction for magnetoresistance. Insert shows the linear coefficient $a_1$ in Eq. $\rho(T) = a_0 + a_1 T$ as a function of $T_c$ for Lu$_2$Fe$_3$Si$_5$ (1), Fe-based system Ba(Fe$_{1-x}$Co$_x$)$_2$As$_2$ (2), and Cu-based system Tl$_2$Ba$_2$CuO$_{6+\delta}$ (3) [31].

The Hall coefficient $R_H$ for the initial polycrystalline sample is relatively small and slightly temperature-dependent, which is in agreement with the measurements of $R_H$ on single crystals, as well as the data on the Fermi surfaces calculated for Lu$_2$Fe$_3$Si$_5$ by the FLAPW method. The Fermi surface consists of two holelike bands and one electronlike band [32], so that the hole and electronic contributions to the Hall coefficient are almost compensated. The irradiation does not lead to a considerable change in $R_H$ (Fig. 4), which serves a kind of evidence that there are no essential doping effects due to the disordering induced by the fast-neutron irradiation.

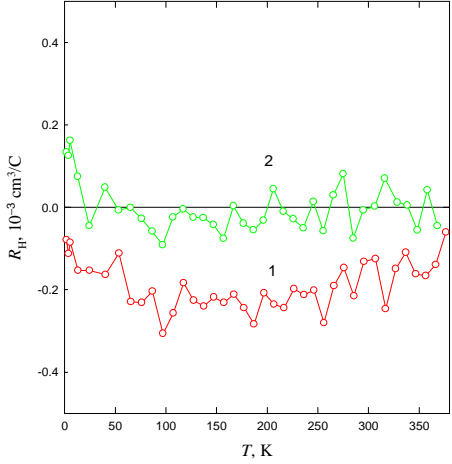

Fig. 4. (color online) Temperature dependence of the Hall coefficient $R_H$ for the $Lu_2Fe_3Si_5$ polycristaline sample: .initial state (1) and irradiated and annealed at 75 $^o$C (2).

Fig. 5 sums the results of annealing of the polycrystalline and single-crystal samples. The reduced resistivity $\rho_0/\rho_{300}$, which is a good measure of the electron mean-free path in $Lu_2Fe_3Si_5$ [28], shows a similar behavior in the single- and poly-crystals as a function of the annealing temperature $T_{ann}$. The intensive recovery of $\rho_0/\rho_{300}$ begins at $T_{ann} \geq 600^o$C only; the radiation defects still survive at relatively high annealing temperatures $T_{ann} \sim 900^o$C.

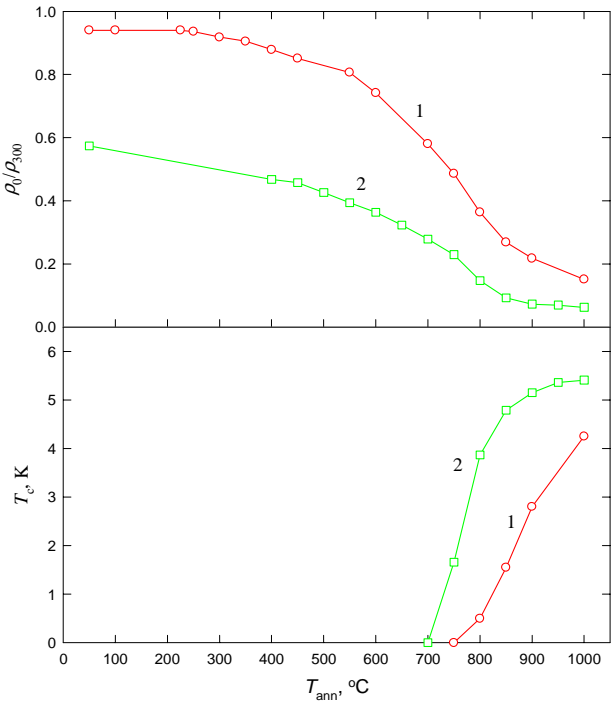

Fig. 5. (color online) Reduced resistivity $\rho_0/\rho_{300}$ and $T_c$ as a function of annealing temperature $T_{ann}$ for polycrystalline (1) and single-crystal (2) $Lu_2Fe_3Si_5$ samples, irradiated to fast neutron fluence $\Phi = 2 \cdot 10^{19}$ cm$^{-2}$ and $\Phi = 5 \cdot 10^{18}$ cm$^{-2}$, respectively.

To compare the suppression of the superconductivity under irradiation with the results of doping with non-magnetic impurities [33, 34, 35], we have drawn $T_c$ determined at 0.5 the normal-state resistivity as a function of the reduced resistivity $\rho_0/\rho_{300}$ (Fig. 6) which does not depend on the sample quality (poly- or single crystal). With increasing $\rho_0/\rho_{300}$, the $T_c$ value is seen to decrease similarly in both the irradiated and doped samples; it goes to zero at $\rho_0/\rho_{300} \approx 0.3$, which corresponds to $\rho_0 \approx 80$ and $\approx 40$ $\mu\Omega$cm for the polycrystalline and single-crystal samples, respectively. The uniform dependence of $T_c$ on $\rho_0/\rho_{300}$ indicates that the only cause of the $T_c$ decrease both under irradiation and doping is the appearance of scattering centers.

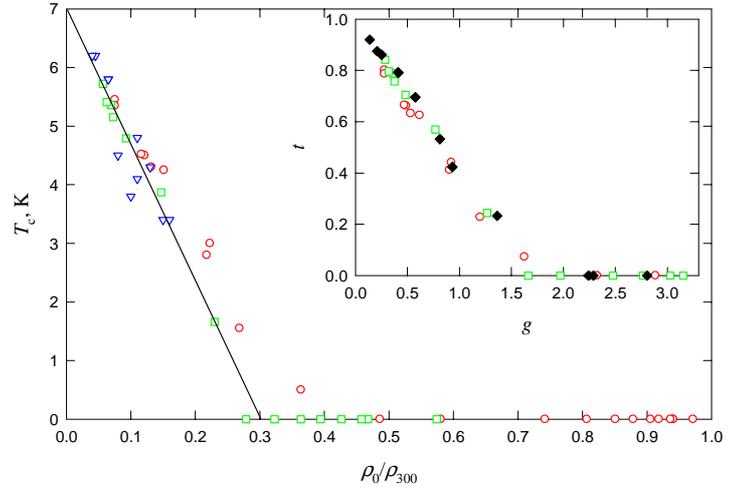

Fig. 6. (color online) $T_c$ as a function of $\rho_0/\rho_{300}$ for the irradiated and annealed polycrystalline (○) and single-crystal (□) $Lu_2Fe_3Si_5$ samples and (Lu-R)$_2$Fe$_3$Si$_5$ (R=Y, Sc, Dy) single crystals (▽) [28]; straight line is drawn by eye. Insert shows $t = T_c/T_{c0}$ vs. $g = \hbar/(2\pi k_B T_{c0}\tau)$ for the polycrystalline (○) and single-crystal (□) $Lu_2Fe_3Si_5$ samples and irradiated La(O-F)FeAs sample (◆) [19].

For comparison with the theoretical models, we made use of the universal Abrikosov-Gor'kov (AG) equation describing the superconductivity suppression by magnetic impurities for the case of $s$-pairing, and by nonmagnetic impurities (defects) for the case of $d$- and $s^\pm$-pairing [36, 37, 38]:

$$\ln(1/t) = \psi(g/t + 1/2) - \psi(1/2), \quad (1)$$

where $g = \hbar/(2\pi k_B T_{c0}\tau) = \xi_0/l$, $\psi$ is the digamma function, $t = T_c/T_{c0}$, $T_{c0}$ and $T_c$ are the superconducting temperatures of the initial and disordered systems, respectively, $\tau$ is the electronic relaxation time, $\xi_0 = (\hbar v_F)/(2\pi k_B T_{c0})$ is the coherent length, $l$ is the mean free path. Equation (1) describes the decrease of $T_c$ as a function of the inverse relaxation time $\tau^{-1}$; superconductivity is suppressed at $g > g_c = 0.28$. The dimensionless parameter $g$ can be constructed from the

experimental values:

$$g = (\hbar\rho_{0ab})/(2\pi k_B T_c \mu_0 \lambda_c^2), \quad (2)$$

where $\lambda$ is the superconducting penetration depth, $\lambda_c = 0.2$ μm [39] for the initial sample in the $H \parallel c$ orientation, $\rho_{0ab}$ corresponds to the $J \parallel ab$ orientation.

The insert in Fig. 6 shows the $t = T_c/T_{c0}$-vs-$g$ curves calculated using Eq. 2 with $\rho_{0ab} = 4\rho_{0c}$ [28] for the single cryatals and the percolation relationship $\rho_{0ab} = 4\rho_0/2.7$ for the policrystals. The quantity $t$ goes to zero at $g \approx 1.5$ which is 5 times as large as the AG value $g_c = 0.28$. A very similar decrease of $t$ versus $g$ was found in the neutron irradiated novel Fe-based compound La(O-F)FeAs (Fig. 6) [19], α-irradiated Nd(O-F)FeAs [20], and proton irradiated Ba(Fe$_{1-x}$Co$_x$)$_2$As$_2$ [21].

There are a number of uncertainties in the identification of the scattering centers contributing to the resistivity $\rho_0$. In the $s^\pm$-pairing model, Eq. 1, only the interband scattering at nonmagnetic impurities is taken into acconunt which is not easily separated from the other contributions (intraband scattering, magnetic scattering etc.). Nevertheless, the AG model significantly overestimates the $T_c$ decrerase in Lu$_2$Fe$_3$Si$_5$ and, probably, in other Fe-based superconductors.

Fig. 7 shows the slope of the upper critical field $-dH_{c2}/dT$ determined at 0.9 the normal-state resistivity, as a function of $T_c$ for the irradiated and annealed Lu$_2$Fe$_3$Si$_5$ polycrystalline and single-crystal samples. The relationship $(-dH_{c2}/dT)_c \approx 2(-dH_{c2}/dT)_{ab}$ holds well for all superconducting crystals, which evidences that the topology of the Fermi surface is not significantly changed by irradiation.

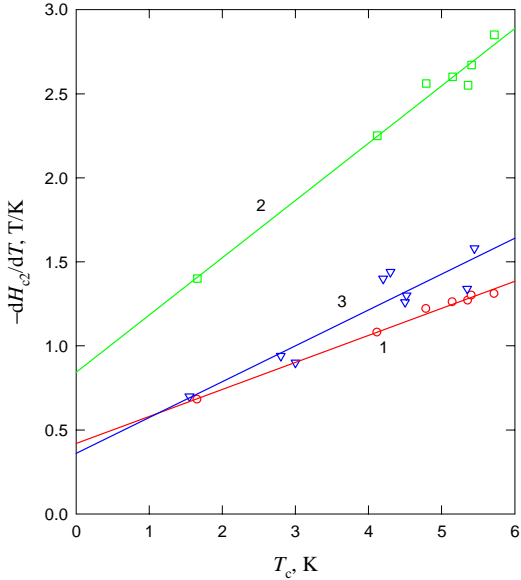

Fig. 7. (color online) The slope of the upper critical field $-dH_{c2}/dT$ as a function of $T_c$ for the irradiated and annealed Lu$_2$Fe$_3$Si$_5$ samples. (1) and (2): single crystal, $H$ is parallel to the $ab$ and $c$ directions, respecrively; (3): policrystalline sample.

The observed behavior can be roughly approximated by a linear dependence. The similar behavior observed in many FeAs-based compounds was attributed to the AG gapless state [40]. It is worth mentioning that $-dH_{c2}/dT \sim T_c$ is also predicted for the isotropic s-wave materials in the clean limit, while in the dirty limit the opposite dependence $-dH_{c2}/dT \sim \rho_0$ (that is the increase in the $-dH_{c2}/dT$ upon decreasing $T_c$) takes place.

The estimation of $g = \xi_0/l$ has shown (Fig. 6) that the superconducting samples belong to the clean ($\xi_0 \ll l$) or the intermediate ($\xi_0 \approx l$) limit for the samples with $T_c \approx 5$ K or $T_c \approx 1.5$ K, respectively. In the expression for the slope of the upper critical field $-dH_{c2}/dT = \phi_0/(0.69 \cdot 2\pi\xi^2 T_c)$ the coherent length $\xi$ can be written in the intermediate limit as

$$1/\xi^2 \approx 1/\xi_0(1/\xi_0 + 1/l), \quad (3)$$

and, hence, $dH_{c2}/dT$ depends on $T_c$ and $\rho_0$ as

$$-dH_{c2}/dT \approx c_1 T_c + c_2 \rho_0. \quad (4)$$

Fig. 8 shows such dependence in the coordinates $(-dH_{c2}/dT)/T_c$ as a function of $\rho_0/T_c$. In the limit $\rho_0/T_c \to 0$ the intercept gives the coherent lenght $\xi = \xi_0$ in the clean limit. According to Eqs. 3, 4, at the doubled intersect we get $\xi = l$ (vertical line in Fig. 8).

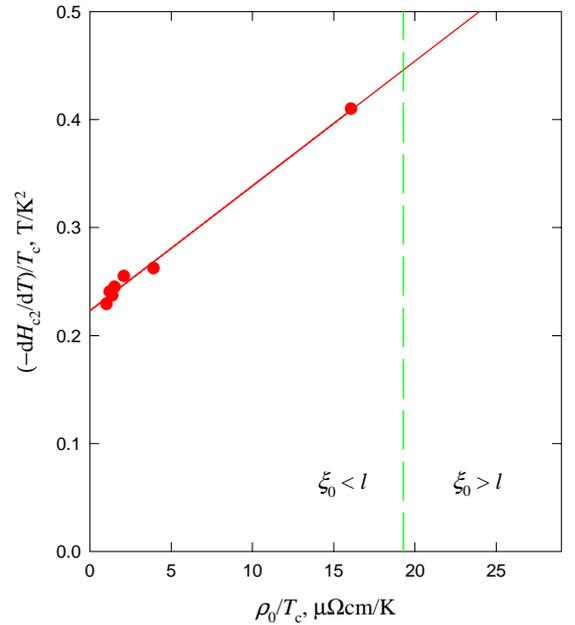

Fig. 8. (color online) $(-dH_{c2}/dT)/T_c$ ($H \parallel ab$) as a function of $\rho_0/T_c$ for the irradiated and annealed Lu$_2$Fe$_3$Si$_5$ single crystal. Straight line is the mean square fitting.

According to Fig. 8, the ration $\xi_0/l = g = 1$ in the single-crystal sample corresponds to $\rho_0 \sim 30$ μΩcm, $T_c \sim 1.5$ K. This is in a satisfactory agreement with the estimation of $g$ according to Eq. 2; $g = 1$ corresponds to $\rho_0 \sim 25$ μΩcm, $T_c \sim 2$ K.

Thus, our estimations of the relation between $\xi_0$ and $l$ clearly show that the initial Lu$_2$Fe$_3$Si$_5$ samples with $T_c \approx 5$ K are ascribed to the clean limit $\xi_0 \ll l$. On the other hand, these estimations are in a visual disagreement with the relation between the anisotropy of resistivities $\rho_{ab}/\rho_c$ and slopes of the upper critical field $(dH_{c2}/dT)_c/(dH_{c2}/dT)_{ab} = \xi_{ab}/\xi_c$. In the clean limit $\xi_{ab,c} \approx (\xi_0)_{ab,c}$ and, hence, the anisotropy of the upper critical field is proportional to $\rho_{ab}/\rho_c$, while in the dirty limit, $\xi_{ab,c} \approx [(\xi_0 l)_{ab,c}]^{1/2}$ and the anisotropy of the upper critical field is proportional to $(\rho_{ab}/\rho_c)^{1/2}$. As it was mention above, the anisotropy of the upper critical field in the initial, irradiated, and annealed single crystal is close to 2, while the anisotropy of resistivities $\rho_{ab}/\rho_c \approx 4$ [28].

Returning to the correlation between the linear term in resistivity $a_1$ and $T_c$ (Fig. 3) it is worth mentioning that the relation $T_c \sim a_1$ is an attribute of many unconventional superconductors. Although the $T_c$ vs. $a_1$ correlation does not seem evident in our case of the irradiated Lu$_2$Fe$_3$Si$_5$ since the $a_1$ term in the low-$T$ resistivity is significantly masked by the logarithmic term for the samples with $T_c \leq 3$ K, the scale of the $T_c$ vs. $a_1$ correlation is very close to that observed for the Fe- and Cu-based high-$T_c$ superconductors (Fig. 3). The linear-in-$T$ resistivity can be explained by the existence of two-dimensional AF spin fluctuations in the theory of nearly AF metals [41, 42]. The AF fluctuations are enhanced significantly near the AF phase in optimally doped high-$T_c$ superconductors, where the temperature dependence of the resistivity changes from the $T^2$- to $T$- law. In Lu$_2$Fe$_3$Si$_5$ with the lower $T_c$ the linear term is meaningful only at low $T < 10$ K (Fig. 3), while at higher $10 < T < 70$ K, the $T^2$-term predominates (Fig. 2).

In conclusion, our results show the fast decrease of the superconducting temperature $T_c$ in the Lu$_2$Fe$_3$Si$_5$ samples under the fast-neutron irradiation. The uniform dependence of $T_c$ on the residual resistivity $\rho_0$ for the case of both irradiation and doping evidences that the decrease in $T_c$ is due to the presence of nonmagnetic scattering centers. The slow changes in the Hall coefficient $R_H$ and $(dH_{c2}/dT)_c/(dH_{c2}/dT)_{ab}$ show that there are no substantial changes in the topology of the Fermi surface caused by irradiation.

The superconductivity disappears when the coherent length $\xi_0$ becomes larger than the mean free path $l$, $g = \xi_0/l > 1$. Such behavior is very similar to that observed in FeAs-based superconductors, but the $T_c$ decrease is ~5 times as slow as that predicted by based on the Abrikosov-Gor'kov equation which describes the superconductivity suppression by nonmagnetic impurities (defects) for the case of $d$- and $s^{\pm}$-pairing.

Our estimations show that the observed correlation of $T_c$ with the slope of the upper critical field $-dH_{c2}/dT$ in the irradiated polycrystalline and single-crystal Lu$_2$Fe$_3$Si$_5$ samples (and probably, in other Fe-based superconfuctors) has a trivial origin: the superconducting samples belong to the clean $\xi_0 \ll l$ (at worst, to the intermediate $\xi_0 \approx l$) limit.

The observed correlation of the liner term $a_1$ in the resistivity $\rho(T) = a_0 + a_1 T$ with $T_c$ testify to the significance of spin fluctuations in the formation of the superconducting state in Lu$_2$Fe$_3$Si$_5$.

This work was carried out with the partial support of the Program of Basic Research of the Presidium of RAS "Condensed Matter Quantum Physics" (Project No. 09-П-2-1005 UB RAS).